 \def\aa{\alpha}

 \def\lb{\left(}
 \def\rb{\right)}

 \def\bb{\beta}

  \def\tw{{t_w+t}}
  \def\te{{t_{\rm erg}}}
\documentstyle[preprint,prl,aps]{revtex}
\begin{document}
%\draft
\title{ \bf AGING ON PARISI'S TREE}

\author{J-P. Bouchaud and D.S. Dean}

\address{Service de Physique de l'Etat Condens\'e}
\address{Direction des Recherches sur l'Etat Condens\'e, les Atomes et les
Molecules}
\address{Commissariat \`a l'Energie Atomique, Orme des Merisiers}
\address{91191 Gif-sur-Yvette CEDEX, France}
\date{\today}
\maketitle

\begin{abstract}
{We present a detailed study of simple `tree' models for off equilibrium
dynamics and aging in glassy systems. The simplest tree describes the landscape
of a random energy model, whereas multifurcating trees occur in the solution of
the Sherrington-Kirkpatrick model. An important ingredient taken from these
models is the exponential distribution of deep free-energies, which translate
into a power-law distribution of the residence time within metastable
`valleys'. These power law distributions have infinite mean in the spin-glass
phase and this leads to the aging phenomenon. To each level of the tree are
associated an overlap and the exponent of the time distribution. We solve these
models for a finite (but arbitrary) number of levels and show that a two level
tree accounts very well for many experimental observations (thermoremanent
magnetisation, a.c susceptibility, second noise spectrum....).

We introduce the idea that the deepest levels of the tree correspond to
equilibrium dynamics whereas the upper levels correspond to aging. Temperature
cycling experiments suggest that the borderline between the two is temperature
dependent. The spin-glass transition corresponds to the temperature at which
the uppermost level is put out of equilibrium but is subsequently followed by a
sequence of (dynamical) phase transitions corresponding to non equilibrium
dynamics within deeper and deeper levels. We tentatively try to relate this
`tree' picture to the real space `droplet' model, and speculate on how the
final description of spin-glasses might look like.

}

\end{abstract}

\vfill


\newpage
%\widetext
\narrowtext

{\it Des journ\'ees enti\`eres dans les arbres. (M. Duras)}

\section { INTRODUCTION}
Aging experiments in glasses and spin-glasses
\cite{Struik},\cite{exp1},\cite{exp2} are now the focus of an intense
theoretical \cite{Russes},\cite{CuKu},\cite{MF},\cite{CuKu2},\cite{CuKuPa} and
numerical \cite{Ander},\cite{rieger},\cite{Ritort} activity. Inspired by
mean-field solutions of the spin-glass problem \cite{MP},\cite{REM}, a simple
picture for the dynamics was proposed in \cite{JPB}, based on the idea that
metastable states act as traps in the phase-space with broadly distributed
trapping times. This picture naturally leads to aging (i.e. non stationary
dynamics) and suggests phenomenological laws for the decay of the
thermoremanent magnetisation (and of the a.c. susceptibility) which are in
quite good agreement with experimental data. \par
However, there are experimental features which are not compatible with the trap
model proposed originally, which in fact corresponds to the phase space of the
simplest kind -- that of the `Random Energy Model' (REM), for which Parisi's
$q(x)$ order parameter is simply $q(x< \overline{x})=0$ and $q(x >
\overline{x})=1$ \cite{GM}. In that model, a metastable state is a single
configuration; hence all dynamics is frozen if the system cannot ``hop out''.
Stated differently, there is no `bottom of the traps' dynamics (see Fig 1-a).

This has various unsatisfactory consequences. For example, the equilibrium a.c.
susceptibility is zero for all frequencies in this model, in plain
contradiction with experiments. This can easily be cured by allowing fast,
small scale fluctuations to reduce the `self-overlap' $q(x > \overline{x})$
from $1$ to a smaller value $q_{EA}$ (the Edwards-Anderson order parameter).
This `dressed' REM behaviour was found recently in simple models for glasses
\cite{DREM}, where the `traps' can be very clearly identified, in particular in
numerical simulations \cite{Krauth}.  \par
More importantly, a REM landscape is insufficient to account for the subtle
effects induced by small temperature cycling \cite{exp3}. Let us mention in
particular the striking memory effect observed on the imaginary part of the
a.c. susceptibility when the temperature is cycled as $T \longrightarrow
T-\Delta T \longrightarrow T$.  The signal rises strongly when the temperature
is first decreased, showing that new dynamical processes are restarted.
However, when the temperature is raised again, the signal recovers exactly the
value it had before the period at $T- \Delta T$, as if this period had not
existed. A possible interpretation \cite{exp3} is that, in our language, there
are `traps within traps' (see Fig. 1-b): the intermediate period corresponds to
non-equilibrium dynamics within a trap -- with hops between `supertraps' frozen
out.

The aim of this paper is thus to analyse in detail and generalize the REM-like
trap model of ref. \cite{JPB} (corresponding to a `one-step' replica symmetry
breaking (RSB) scheme \cite{GM}) to a fully foliated tree structure (`full
replica symmetry breaking' \cite{MP}). A large amount of papers already studied
various types of dynamics on trees \cite{treemodels}, directly inspired from
Parisi's ultrametric construction. We however believe that our model is closer
both to Parisi's construction and to reality (in particular because most of
these studies analyse deterministic trees). As in its REM version, aging
appears most naturally within this framework and provides an excellent fit of
experimental data, which in turn allow to determine the structure of the $q(x)$
characterizing the tree. We show that the data is already very well accounted
for within a `two-step' RSB approximation scheme.
 We discuss qualitatively $T-$jump experiments and `noise second spectrum'
\cite{Weiss} within our model.  We speculate on how this `tree of states' could
be
interpreted in finite dimensional space, and rephrase the droplet model of
Fisher and Huse \cite{FH} in that context. We suggest that there is a
spin-glass transition temperature associated with each length scale. We believe
that our model, although
still phenomenological, is of help to grasp the subtleties involved in the
non-equilibrium dynamics of random or glassy systems.
 It also sheds light on the
analytical results recently obtained for `toy' microscopic spin-glass models
exhibiting replica symmetry breaking \cite{CuKu2},\cite{MF}.

\section { THE MODEL AND SOME ANALYTIC RESULTS}

\subsection {  The Single Layer Tree (REM)}

The simplest version of the model, already considered in \cite{JPB}, is defined
as follows: the $N$ possible `states' of the system live on the leaves  of the
one layer tree drawn in Fig. 1-a. To each branch $\alpha$ is associated an
energy barrier $\epsilon_\alpha$ distributed as (minus) the energy in the REM,
i.e.:
$$ \rho(\epsilon) =  {1 \over T_g} \exp [-{\epsilon \over T_g}] \eqno(1)$$
where $T_g$ is the glass transition temperature. To each $\alpha$ thus
corresponds  an Arrhenius hopping time $\tau_\alpha \equiv \tau_0 \exp
{\epsilon_\alpha \over T}$, where $\tau_0$ is a microscopic time scale. From
Eq. (1), one finds that the $\tau_\alpha$ are (quenched) random variables
distributed according to a power-law of index $x = {T \over T_g}$:
$$ p(\tau) = {x \tau_0^x \over \tau^{1+x}} \ \ \ \ \ (\tau \geq \tau_0)
\eqno(2)$$
We shall call $\tau$ an [x]-variable. For $0 <x \leq 1$ (i.e. for $T<T_g$), the
mean value of an [x]-variable is infinite: this will lead to the aging
phenomenon \cite{JPB}. \par
The model is further defined by saying that hopping rate from state $\alpha$ to
state $\beta$ is $1 \over N \tau_\alpha$. This simple dynamics has the virtue
of giving the correct Boltzmann equilibrium distribution : for finite $N$, the
equilibrium weight is: $P_{\rm eq}^\alpha = Z^{-1} \exp {\epsilon_\alpha\over
T}$. The overlap between two states $\alpha,\beta$ is $q_0 \equiv 0$, while the
self-overlap is taken to be $q_{EA} \leq 1$, independently of $\alpha$. The
`spin-spin' correlation function $C(t_w+t,t_w) \equiv \langle S(t+t_w)S(t_w)
\rangle$ within this model is thus naturally defined as:
$$ C(t_w+t,t_w) \equiv q_{EA} \Pi_1(t,t_w) \eqno(3) $$
where $\Pi_1(t,t_w)$ is the probability that the process has not jumped between
$t_w$ and $t_w+t$ : this is the basic object we shall proceed to calculate.

Note that from Eq. (3), $C(t_w,t_w)$ can be smaller than $1$. Strictly
speaking, this is true in the limit $\tau_0 =0$: it means that there is some
dynamics taking place at short times, and reflecting the fact that a `state'
$\alpha$ is a ensemble of configurations mutually accessible within microscopic
time scales. One could thus add branches connecting all these configurations to
the `node' $\alpha$ (see Fig 1-a); the `trapping' times associated with this
level of the tree are however such that their average is finite. A natural way
is to continue Parisi's $q(x)$ function for $x>1$ and $q_{EA} < q < 1$
\cite{KuPaVi}, such that the corresponding $\langle \tau \rangle = {x \over
1-x} \tau_0$ are microscopic. This corresponds to the {\it equilibrium} part of
dynamics (see \cite{Som}), while $C(t,t_w)$ corresponds to the non-stationary
(aging) part. This important idea will be generalized below: for a finite
number $N$ of branches, or an upper cut-off in the distribution (2),
the distinction between equilibrium and  non-equilibrium becomes itself
time-dependent.

The system is assumed to start in a randomly chosen initial metastable state.
Rather than work at fixed $t_w$, it is convenient to work in Laplace transform,
and define $\hat \Pi_1(t,E)= \int_0^\infty E dt_w \ \exp (-Et_w) \
\Pi_1(t,t_w)$. This is because, conditional on the value $\tau_\alpha$, the
amount of time spent in a trap is exponentially distributed -- using some
standard properties of exponential random variables then makes calculations
relatively straightforward.
\par We define $p(E,\tau_\alpha)$ to be the probability that at a random
exponential time $\hat t_w$ of rate  $E$ the system is found in  trap $\aa$
with hopping rate
${1\over \tau_\alpha}$. To be in trap $\alpha$ at time $\hat t_w$, one
possibility is that the system starts in trap $\alpha$ and time $\hat t_w$
occurs before the system escapes from that trap. Clearly one starts in trap
$\aa$ with probability
${1\over N}$, the probability that $\hat t_w$ occurs before escaping trap $\aa$
is then given by property (3) of exponential random times as $ E/(E + {1\over
\tau_\aa})$. However, to start with, the system may be in any of the  traps
$\bb \quad 1\leq \bb\leq N$
but escape from it before time $\hat t_w$. Conditional on this having happened,
the memoryless property of the exponential time (property (A-1)) means that the
probability
of being in trap $\aa$ at time $\hat t_w$ is just $p(E,\tau_\aa)$, that is the
game starts anew. Putting all of this together allows us to write down the
renewal equation
$$ p(E, \tau_\aa) = {1\over N} {E\tau_\aa \over E\tau_\aa +1} +{1\over N}
\sum_{\bb} ^N
{1\over E\tau_\bb +1} p(E,\tau_\aa) \eqno(4) $$
hence
$$ p(E, \tau_\aa) = {{\textstyle{E\tau_\aa \over E\tau_\aa +1}}\over
\sum_{\bb} ^N
{E\tau_\bb\over E\tau_\bb +1}} \eqno(5)$$
In the case where $N$ is large we may use the approximation
$$  p(E, \tau_\aa) = {{E\tau_\aa \over E\tau_\aa +1}\over N\langle
{E\tau_\bb\over E\tau_\bb +1}\rangle}\eqno(6)$$
the angled brackets indicating averaging over the disorder. In the regime where
$E\tau_0 \ll 1$ (equivalent to $t_w \gg \tau_0$) we obtain
$$ \langle
{E\tau_\bb\over E\tau_\bb +1}\rangle \sim x\tau_0^x E^x c(x) \eqno(7)$$
where
$$ c(x) = \Gamma(x) \Gamma(1-x) = {\pi \over \sin(\pi x)}$$
We now wish to calculate the probability that after an initial waiting time
$t_w$ the system has not decayed from the state it was in at time $t_w$ after a
subsequent time $t$ has elapsed after $t_w$. Keeping the waiting time
exponential, we find
$$ \hat \Pi_1(t,E)= \langle \sum_{\bb} ^N p(E, \tau_{\bb}) e^{-{t\over
\tau_\bb}}\rangle
\sim   {1\over E^x c(x)}\int_{\tau_0 }^\infty d\tau \tau^{-x-1} {E\tau\over
E\tau +1} e^{-{t\over \tau}}\eqno(8) $$
Now, making a judicious change of variables and noting that $ \exp (-Ew) \equiv
\int_0^\infty Edv \exp (-Ev)$ $\theta(v-w)$, one finally finds:
$$ \Pi_1(t, t_w) = {\sin(\pi x)\over \pi}\int_{{t\over t + t_w}} ^1
du(1-u)^{x-1} u^{-x} \eqno(9) $$
This last form shows very clearly that the correlation function only depends,
within this model, on the rescaled time ${t \over t+t_w}$, i.e $t$ over the
`age' of the system $t+t_w$: as argued in \cite{JPB}, only traps such that
$\tau \sim O(t_w)$ have an appreciable probability to be observed after time
$t_w$. From the same argument, the average energy decreases, in this model, as
${\cal E}(t_w) - {\cal E}(t=0) \sim  - T \log({t_w \over \tau_0})$.

Two asymptotic regimes are of interest. If $t \ll t_w$ one obtains
$$ \Pi_1(t,t_w) \sim 1 - {\sin(\pi x) \over \pi (1-x)}\lb{t\over t +
t_w}\rb^{1-x}\eqno(10-a) $$
In the regime $t \gg t_w$, we obtain
$$ \Pi_1(t,t_w) \sim  {\sin(\pi x) \over \pi x}\lb{t_w\over t + t_w}\rb^x
\eqno(10-b)$$
$C(t_w+t,t_w)$ has thus the `weak-ergodicity breaking' property
\cite{JPB},\cite{CuKu}:
$$ \lim_{t \longrightarrow \infty}  C(t_w+t,t_w) = 0 \ \ \ \ \ \hbox{ for all
finite}\  t_w \eqno(11-a) $$
but
$$ \lim_{t_w \longrightarrow \infty}  C(t_w+t,t_w) = q_{EA}\ \ \ \ \ \ \hbox{
for all finite}\  t\eqno(11-b) $$
meaning that `true' ergodicity breaking only sets in after infinite waiting
times \cite{Werg}.

The critical case $x=1$ is of special interest, since it leads to aging but
with an `anomalous' scaling (i.e. not as $t \over t_w$), we find:
$$\Pi_1(t,t_w) = \tau_0 \int_0^{t_w \over \tau_0} du{1 - \exp\lb- \ { t_w + t
-u\tau_0 \over \tau_0}\rb \over (\log u) (\tw - u\tau_0)} \eqno(12-a)$$
which leads to
$$\Pi_1(t,t_w) = 1 -{\log({t\over \tau_0}) \over \log({t_w\over \tau_0})}\qquad
{\rm for} \ \log({t\over \tau_0}) \ll \log({t_w\over \tau_0})\eqno(12-b)$$
and
$$\Pi_1(t,t_w) = {t_w \over t \log({t_w\over \tau_0})}\qquad {\rm for} \ t \gg
t_w\eqno(12-c)$$

The above results pertain to the `aging' case $x \leq 1$. For completeness, and
also because they may be relevant to describe the equilibrium dynamics at short
times (or for $T>T_g$, see \cite{Levelut},\cite{KuPaVi}), let us give the
corresponding results for $x > 1$ in the limit $t_w \gg \tau_0$:
$$\Pi_1(t,t_w) = 1 - {x \over x-1} ({t \over \tau_0}) \qquad {\rm for} \ t \ll
\tau_0\eqno(13-a)$$
$$\Pi_1(t,t_w) = (x-1) \Gamma(x) ({\tau_0 \over t})^{x-1}\qquad {\rm for} \
\tau_0 \ll t \ll t_w\eqno(13-b)$$
and
$$\Pi_1(t,t_w) =(x-1)^2 \Gamma(x)({\tau_0^{x-1} t_w \over t^x})\qquad {\rm for}
\ t \gg t_w\eqno(13-c)$$
Note the major difference between these last results and those given in Eq.
(10): for $\tau_0 \longrightarrow 0$, $C(\tw,t_w)$ is identically zero for
$x>1$, but {\it remains non zero} for $x < 1$. In this sense, the crossing of
$x=1$ corresponds to a true dynamical freezing transition.

\subsection {  The Multi-Layer Tree (RSB)}

The above single layer tree model can be generalized, following the
construction of M\'ezard, Parisi and Virasoro for the tree of states (see
\cite{MP}, and also \cite{Parisi}).
The hopping rate from state $\alpha$ to state $\beta$ is set by a random escape
time $\tau_{\alpha,\beta}$ the statistics of which depends on the overlap
$q_{\alpha,\beta}$ between states. In Parisi's tree, the free-energies
associated to a certain level are distributed according to Eq. (1), but with a
level-dependent $T_g$, and hence with the overlap associated with that level.
$\tau_{\alpha,\beta}$ is thus an [x]-variable, with a $q-$dependent $x$ which
is precisely the inverse of Parisi's order parameter function $q(x)$. In the
case of the REM ($q(x< \overline{x})=0$ and $q(x > \overline{x})=q_{EA}$) we
recover the single-layer tree since all the $\tau$'s have the same statistics.
\par
The procedure we follow to solve this model for a finite number $M$ of layers
(finite RSB) is a generalisation of the one presented above. We introduce the
probabilities $\Pi_{j}(t,t_w)$, $j=1,...,M$ that the process has never jumped
beyond the $j^{\rm th}$ layer of the tree between $t_w$ and $t_w+t$. $j=M$
corresponds to the deepest level of the tree and $j=1$ the upper level of the
tree -- see Fig 1-b -- and hence $\Pi_1$ is the same as the one introduced for
the single level tree. Introducing the $j^{\rm th}$ level overlap $q_j$ (with
$q_0=0$ and $q_M=q_{EA}$), the spin-spin correlation function is clearly:
$$
C(t_w+t,t_w) = \sum_{j=0}^{M}  q_j \left[\Pi_j(t,t_w) - \Pi_{j+1}(t,t_w)\right]
\eqno(14)
$$
with the convention that $\Pi_0(t,t_w)=1$ and $\Pi_{M+1}(t,t_w)=0$ (all
processes inside a state have indeed taken place on the time scales $\gg
\tau_0$ - see the discussion after Eq. (3).)
The precise construction of the multi-layer tree is a follows, the $(j+1)$th
layer tree is constructed by adding $N_{j+1}$ branches at the end of each
branch of the $j$ layer tree -- the new states are now at the ends of these new
branches. Associated with each new branch $\aa$ is a new hopping rate ${1\over
\tau_{\aa({j+1})}}$ with which the process hops back to level $j$ and then
falls back into the $(j+1)$th layer. For simplicity, we define this hopping to
be independent of all other hopping on the tree, i.e. the hopping dynamics is
in parallel. We therefore have transitions from states characterised by inverse
hopping rates $\{\tau_1,\tau_2,\cdots,\tau_{k-1},\tau_k,\cdots
,\tau_{j-1},\tau_j\}$ to states
$\{\tau_1,\tau_2,\cdots,\tau_{k-1},\tau^\prime_k ,\cdots,
\tau^\prime_{j-1},\tau^\prime_j\}$ at rate ${1\over \tau_k}$ (the primes
denoting new rates chosen within the tree structure) and all these transitions
are in parallel.
 The $\tau_j$ are quenched random variables chosen according to the law
$$ p(\tau_j) = {x_j \tau_{0j}^{x_j} \over \tau_j^{1+x_j}} \ \ \ \ \ (\tau_j
\geq \tau_{0j})\eqno(2')$$
where $0 <x_1 < x_2 \cdots x_{j-1} < x_j <1$. These times $\tau_j$ corresponds
to (free) energy barriers $\epsilon_j$ distributed as in Eq. (1), but with a
level dependent freezing temperature, $T_{gj}$. Hence, again, the dynamics of
this model reproduces, at long times and for a finite number of branches, the
Boltzmann equilibrium: the total  weight of a branch at level $j$ is given by
$Z_j^{-1} \exp {\epsilon_j \over T}$.

For technical reasons we shall also assume that microscopic time scales
associated with levels lower down the tree are much smaller than those
associated with microscopic time scales further up the tree -- this is also a
reasonable hypothesis from the physical point of view. Therefore we shall
impose $\tau_{0{j+1}} \ll \tau_{0j}$.

Once again we work with an exponential random time $\hat t_w$ of rate $E$. We
define $p_j(E;\tau_1,\cdots,\tau_j)$ to be the  probability of finding the
system in a set of traps characterised by  $\{\tau_1,\cdots,\tau_j\}$ at time
$\hat t_w$. We calculate the $p_j$ inductively as follows
$$ p_{j+1}(E; \tau_1,\cdots,\tau_{j+1}) =  p_{j}(E; \tau_1,\cdots,\tau_{j}
)\cdot p(E_{j+1},\tau_{j+1})\eqno(15) $$
where $E_j = E + \sum_{i = 1}^{j-1} 1/\tau_i$ for $j\geq 2$, $E_1 = E$ and $p$
is the result for the single layer tree. This comes from an application of
Bayes' theorem and using properties (A-2) and (A-3) of exponential times.
Consequently, $ p_j(E; \tau_1,\cdots,\tau_{j}) = \prod_{i=1}^j p(E_i,\tau_i)$.
The fact that we have chosen $\tau_{0j} \ll \tau_{0(j-1)}$, and in addition
chosen $M$ to be reasonably small, means that we are assured that $E_j
\tau_{0j} \ll 1$ and hence we may use the small $E$ asymptotic result for the
$p(E_j,\tau_j)$ in the expression above, yielding
$$ p_j(E; \tau_1,\cdots,\tau_{j}) = \prod_{i=1}^n {{E_i\tau_i \over E_i\tau_i +
1}\over N_i x_i c(x_i)E_i^{x_i} \tau_{0i}^{x_i}}\eqno(16) $$
We now use
$$ \hat \Pi_j(t,E) = \sum_{\{\tau_1 \cdots \tau_j\}} p_j(E;
\tau_1,\cdots,\tau_{j} )\exp(-t\sum_{i=1}^j {1\over \tau_i})\eqno(17)$$
and the fact that the $N_i\to \infty$ and hence all combinations of $\{
\tau_1,\cdots,\tau_{j} \}$ exist, allows us to write
$$ \hat\Pi_j(t,E) = \int_{\tau_{01}}^\infty d\tau_1 \cdots
\int_{\tau_{0j}}^\infty d\tau_j\exp\lb -t\sum_{i=1}^j {1\over \tau_i}\rb
\prod_{i=1}^j {{E_i\tau_i^{-x_i} \over E_i\tau_i + 1}\over
c(x_i)E_i^{x_i}}\eqno(18)$$
One again attempts an integration by substitution, starting with $\tau_j$ and
proceeding down to $\tau_1$. On inverting the Laplace transform one obtains
$$ \Pi_j(t,t_w) = {\int^1 _0 \prod_{i = 1} ^j du_i u_i^{-x_i} (1-u_i)^{x_i -1}
\theta(\prod_{i = 1} ^j u_i - {t\over t+ t_w}) \over  \prod_{i = 1}^j
B(x_i,1-x_i)} \eqno(19) $$
where $\theta$ is the Heaviside step function.
The above formula has an immediate probabilistic interpretation. If $U_i$ are
independent random variables from  Beta distributions of indices $1- x_i$ and
$x_i$ respectively, then
$$ \Pi_j(t,t_w) =  P(\prod_{i = 1} ^j U_i > {t\over t+ t_w}) \eqno(20) $$
A useful rearrangement of equation (19) for asymptotic analyis is
$$  \Pi_{j+1}(t,t_w) = \Pi_j(t,t_w) - {\int^1 _0 \prod_{1=1} ^j du_i u_i^{-x_i}
(1-u_i)^{x_i -1}
\theta(\prod_{i = 1} ^j u_i - s)B_{{s\over\Pi_{1}^j u_i}}(1-x_{j+1},
x_{j+1})\over \prod_{i=1}^{j+1}B(x_i,1-x_i)} \eqno(21) $$
where $s = {t\over t+t_w}$ and
$$ B_\gamma (x,y) \equiv \int_0 ^\gamma du u^{x-1} (1-u)^{y-1}\eqno(22) $$
is the incomplete Beta function (clearly $B_1(x,y) \equiv B(x,y)$).
The above rearrangement also makes clear that the obvious inequality
$\Pi_{j+1}(t,t_w) <  \Pi_j(t,t_w)$ is satisfied.
\par
Let us focus on the short time and long time asymptotics implied by the above
analytical forms.  For $t \ll t_w$, we find that
$$
C(t_w+t,t_w) = q_{EA} -  (q_{EA}-q_{M-1}) {\sin(\pi x_M) \over \pi (1-x_M )}
\prod_{j=1}^{M-1} {B(x_j,x_M-x_j) \over B(x_j,1-x_j)} \lb{t\over t +
t_w}\rb^{1-x_M}+$$
$$
+O(({t\over t_w})^{1-x_{M-1}}) \eqno(22)
$$
This form shows that the initial decay is, as expected, dominated by the
deepest level of the tree. The possibility of `jumping' to higher levels during
the
waiting period however acts to reduce the effective waiting time by a factor
$\left[\prod_{j=1}^{M-1} {B(x_j,x_M-x_j) \over B(x_j,1-x_j)}\right]^{-{1 \over
1-x_M}} < 1$. Each of these large scale jumps indeed completely restarts the
small scale dynamics.

In the limit $t \gg t_w$, on the other hand, the asymptotic decay is given by:
$$
C(t_w+t,t_w) = q_{1} {\sin(\pi x_{1}) \over \pi x_{1}}\lb{t_w\over t +
t_w}\rb^{x_{1}} + O(({t_w \over t})^{x_{1}+x_{2}}) \eqno(23)
$$
Note that, again, $C(t_w+t,t_w)$ is only a function of the ratio $t\over t_w$.
This would not be true if the number of branches $N_{j}$ is finite, as we shall
discuss now.

\subsection { Finite number of branches and interrupted aging.}

Let us now discuss the case where the number of `states' $N$ is finite,
starting with the single layer tree model. As can be seen from Eq. (5), or
directly from the argument giving the order of magnitude of the largest $\tau$
drawn from distribution (1), the results established in section II.1 rely on
the inequality:
$$ t+t_w \ll \tau_0 N^{1\over x} \equiv \te \eqno(24) $$
In the opposite case, the dynamics ceases to depend on $t, t_w$, and stationary
dynamics resume and aging is `interrupted': $\te$ is the ergodic
time of the problem. $\Pi_1$ asymptotically behaves as in Eqs. (10), but with
the variable $t \over t_w$ replaced by ${t \over \te}$. The scaling of
$C(\tw,t_w)$ as a function of $t \over t_w$ is thus expected to degrade
progressively as $t,t_w$ approach $\te(N)$. This was proposed as an explanation
for the small (but systematic) deviation from a perfect $t \over t_w$ scaling
observed
experimentally \cite{JPB}. Let us note that very similar results would of
course hold if $N$ is infinite but the power-law distribution (2) happens to be
truncated beyond a certain $\te$.

The case of a multilayer tree is interesting, but rather complex -- we shall
thus only give qualitative arguments. Let us assume that the ergodic time at
level $j$ is now $\te^j$. Is is reasonable to assume that the
hierarchy of the $x_j$ (slower dynamics as one gets higher up the tree) is
maintained for the ergodic times, i.e.: $\te^j \gg
\te^{j+1}$. For a given waiting time $t_w$, a specific level of the tree $j_w$
is naturally selected through the condition

$$
\te^{j(t_w)} \gg t_w \gg \te^{j(t_w)+1} \eqno(25)
$$
meaning that all the levels `below' $j(t_w)$ are equilibrated, while those
above
are still aging. In this case, even if $x_{j(t_w)+1} < 1$, the corresponding
transitions
are equilibrated -- reducing further the effective value of $q_{EA}$,
renormalized by small scales dynamics. The interesting
point, however, is that the effective `ergodic' time for which deviations from
a $t \over t_w$ scaling first become visible is now $\te^{j(t_w)}$ which
becomes,
through Eq. (24), {\it dependent on $t_w$ itself}. [It might well be that the
ergodic time determined experimentally in \cite{JPB} is only the one
corresponding to the particular level of the tree which happens to be probed on
the $1-10000$ seconds time scales].

Another consequence of these nested equilibrium time scales is the following:
if a perturbation is applied to the system at time $t_w$, the response at time
$t_w +t$ will depend on how much the system has been able to reequilibrate
during time $t$. The response is thus expected to depend on the particular
level $j$  defined as in Eq. (25), but with $t$ replacing $t_w$. Hence, for $t
\ll t_w$, one expects that $C(\tw,t_w) = q_{j(t)}$, leading to violations of
the ${t \over t_w}$ scaling at short times too. Said differently, the effective
`microscopic' time scale $\tau_0$ is replaced by $ \te^{j(t_w)+1}$
\cite{deltah}. The existence of different time domains in the plane $t_w,t$ was
in fact suggested in refs. \cite{MF},\cite{CuKu2}.

\section { RELATION WITH MEASURABLE QUANTITIES. }

\subsection {Constant temperature data}

Although noise measurements can be and have been performed on spin-glasses to
access directly the correlation function computed above, the response function
-- measured through the thermoremanent magnetisation (TRM) or the a.c.
susceptibility -- is of much easier access. Most of the available data thus
gives information on the response function $R(t,t') \equiv {\partial <S(t)>
\over \partial H(t')}$ (but see below). For equilibrium dynamics, the response
and correlation functions are related through the Fluctuation-Dissipation
theorem (FDT). For non-stationary dynamics, an extended form of this theorem
has recently been proposed \cite{CuKu},\cite{CuKu2},\cite{MF},\cite{CuKuPa}.
This
relation can be written as:
$$
T \  R(t,t') = \theta(t-t') X[C(t,t')] {\partial C(t,t') \over \partial t'}
\eqno(26-a)
$$
where $X(.)$ is a certain function which can be computed within simple models
\cite{CuKu},\cite{CuKu2},\cite{MF}. For the SK model near $T_g$ and the
toy-model of \cite{MF}, $X$ is the inverse of the function $q(x)$, whereas for
the spherical $p-$spin model, $X$ is a constant $\leq 1$ (depending on
temperature). (Equilibrium dynamics corresponds to $X \equiv 1$ and one
recovers the usual FDT.)

Within the present context, the response function can be computed by assuming
that to each trap $i$ is associated a certain magnetisation $m_i$ with
probability $\rho(m)$, which we shall take to be even $\rho(-m)=\rho(m)$. The
hopping rate from state $i$ to state $j$ is modified in the presence of a field
H by a factor $\exp - \left[{H \{\zeta m_i - (1-\zeta) m_j\} \over T}\right]$
which ensures the correct equilibrium weighting for all values of $\zeta$. It
is then easy to establish that \cite{Silvio}:
$$
T \  R(t,t') = \theta(t-t') \left[(1-\zeta) {\partial C(t,t') \over \partial
t'}
- \zeta  {\partial C(t,t') \over \partial t} \right]
\eqno(26-b)
$$
If $C(t,t') = C(t-t')$, the usual FDT is recovered. If -- as is the case here
-- $C(t,t')=C({t\over t'})$, then one finds that Eq. (26-a) holds with $X = (1
-\zeta) + \zeta {t' \over t} \equiv  (1 -\zeta) + {\zeta \over C^{-1}({t\over
t'})}$. Note that in the early regime corresponding to equilibrium dynamics ($t
= t' + O(\tau_0)$), $X \equiv 1$ independently of $\zeta$.

Now, the TRM $M(t_w+t,t_w)$ is defined as:
$$
M(\tw,t_w) = \int_{0}^{t_w} dt' R(\tw,t') \ \ H \eqno(27)
$$
where $H$ is the applied field between $0$ and $t_w$. Defining the function
$Y(q)$ through $X(q)={dY(q)\over dq}$ and using Eq. (26), we obtain that the
experimentally measured TRM and $C(\tw,t_w)$ are related through \cite{CuKu}:
$$
M(\tw,t_w) =  Y[C(\tw,t_w)] h \eqno(28)
$$
where $h \equiv {H \over T}$. The simplest assumption of a REM with $q(x<
\overline{x})=0$, $q(x > \overline{x})=q_{EA}$ and $X={\rm const.}$ leads to:
$$
M(\tw,t_w) = q_{EA} X h  \Pi_1(t,t_w)\eqno(29)
$$
Experimentally, the decay of $M(\tw,t_w)$ at short and large times is indeed
very well described by the forms given in Eqs. (10-a, 10-b). In particular, the
scaling as a function of the reduced time $t\over t_w$ is reasonably well
obeyed (but see \cite{JPB}-b] and section II.3 above). In fact, the
experimental value of $\overline x$ is in the range $0.6 - 0.9$ if extracted
using the short-time expansion (10-a), and in the range $0.05-0.2$ if extracted
using the long-time expansion (10-b) \cite{JPB}, where it was called $\gamma$].
This suggests that a more complicated RSB scheme is needed, generating a tree
with at least two levels -- allowing for the existence of two independant
exponents $x_1$ and $x_2$. In the spirit of Parisi's step by step construction
of the SK solution, we shall thus assume that $q(x)$ is given by:
$$
q(x < x_1)=0 \qquad q(x_1 < x < x_2)=q \qquad q(x>x_2)=q_{EA} \eqno(30)
$$
In order to obtain $M(\tw,t_w)$ we first assume that $X={\rm const.}$, or
$\zeta=0$. This assumption is natural within our picture  since $q_j$
corresponds to the fraction of `frozen' spins at level $j-1$: these spins thus
retain their magnetisation when the process `jumps' at level $j-1$ leading to:
$$
M(\tw,t_w) \propto  [(q_{EA}-q)  \Pi_2(t,t_w) + q \Pi_1(t,t_w)]  \eqno(31)
$$
This form is only valid after the short time ($x > 1$) relaxation has taken
place; just before the field is cut, the field-cooled magnetisation is by
definition $M(t_w^-,t_w) := \chi_{FC} H$, where $\chi_{FC}$ is roughly
temperature independent. We choose units such that $M(t_w^-,t_w) \equiv 1$. The
four fitting parameters are thus $x_1,x_2,{q\over q_{EA}}$ for the {\it shape}
of the TRM decay and ${q_{EA} X \over T}$ for the $y-$axis scale. We show in
Fig 2-a and 2-b the best fits obtained for both studied samples,
Ag:Mn$_{2.6\%}$ (AgMn) at $T= 9 K$ (0.87 $T_g$) and
CdCr$_{1.7}$In$_{0.3}$S$_{4}$ (CrIn) at $T=10 K$ ($0.6 T_g$), plotted as a
function of the rescaled variable $s ={t \over t+t_w}$. As can be seen, these
fits are excellent on the whole time regime, which extends, for CrIn, from $10$
sec. to $2.3 \times 10^5$ sec \cite{longtime}. The value of the fitting
parameters for different temperatures and samples are given in Table 1 and 2.
Assuming that $X$ is not critical at $T_g$, we can extract values for the
exponents $\beta,\beta'$ defined as $q_{EA} \simeq ({T-T_g \over T_g})^\beta$
and ${q} \simeq ({T-T_g \over T_g})^{\beta'}$: $\beta \simeq 0.65$ and $\beta'
\simeq 1.0$ (both samples give the same values for these exponents, within
$10\%$.) Our value of $\beta$ is compatible with other determinations ($\beta
\simeq 0.5-1$) \cite{BY}.

The fits shown in Fig. 2-a, 2-b implicitely assume that $x_2 < 1$. We have also
tried the case $x_2 = 1$ -- the free parameter now being $\tau_0$. The
resulting fit is of much poorer quality; however, as will be discussed in
section IV (Fig 6 below), the initial part of the TRM decay for different $t_w$
is rather well rescaled when plotted as a function of
${\log({t\over \tau_0}) \over \log({t_w\over \tau_0})}$, as suggested by Eq.
(12-b). We interpret this as a sign that $q(x)$ is in fact continuous, at least
near $x=1$.

We have also tried to fit the data using different values of $\zeta$. The
quality of the fit remains excellent if $\zeta \leq {1 \over 2}$ (although the
values of $q_{EA},q$ are somewhat changed), and deteriorates for larger values
of $\zeta$. One can also use the form of $X(C)$ suggested by the models studied
in \cite{CuKu},\cite{CuKu2},\cite{MF}, i.e., the inverse of the function $q(x)$
defined by Eq. (30). Defining a characteristic time $t^*$ such that
$C(t_w+t^*,t_w)=q$, the TRM takes the following form (for $t=0^+$):
$$
M(\tw,t_w)= [x_1 (C(\tw,t_w)-q) + x_2 q] h \quad (t < t^*)\eqno(32-a)
$$
$$
M(\tw,t_w)= x_2 C(\tw,t_w) h\quad (t>t^*) \eqno(32-b)
$$
i.e., the decay rate is discontinuous at $t=t^*$. We have plotted in Fig 3 the
best fit obtained using Eqs (32): the initial and final part of the curve can
be accounted for, but not the intermediate region around $t=t^*$. This suggests
that either the function $X$ is (in our case) not connected simply to $q(x)$
and
is a constant for the range of correlations probed, or else that a continuous
RSB scheme is needed (to remove any discontinuity in decay rate). More work on
this point is certainly needed, in particular to extend our formulae to a
continuous tree.
\par
Let us now turn to the a.c. susceptibility measurements. The object which is
measured is now $\chi(\omega,t) = \int_{-\infty}^t R(t,t') \exp(i \omega
(t'-t)) dt'$. Let us first assume a REM-like dynamics. Then using Eqs. (26),
and in the limit $t \gg \omega^{-1}$ (actually needed for an accurate
measurement of $\chi(\omega,t)$, we find that:
$$
\chi(\omega,t) = {X q_{EA} \over T} \int_0^{\infty} du \lb \exp [-i(\omega t)u]
-1 \rb {\sin (\pi\overline x) \over \pi} u^{-\overline x} \eqno(33-a)
$$
(the $\omega=0$ value of $\chi'$ has been removed). Defining $\overline \omega
\equiv \omega t$ and changing variables,
$$\chi(\omega,t) \equiv \chi(\overline \omega) = X{q_{EA}\over T} {\sin
(\pi\overline x)\over \pi} \Gamma(1-\overline x) e^{-i \delta}  {\overline
\omega}^{\overline x -1}
\eqno(33-b)
$$
in agreement with the arguments given in \cite{JPB}. Eq. (33-b)
shows that:
\par
(i) the relevant variable is again the rescaled time $\omega t$: this is very
well obeyed by experimental data.
\par
(ii) There is a `constant loss angle' $\delta = {\pi \over 2} (1-\overline x)$
between the aging part of $\chi'$ and $\chi''$, as expected from Kramers-Kronig
relations.\par
(iii) For a multilayer tree, the same formula (33-b) will hold in the limit
$\overline \omega \gg 1$, with $\overline x$ replaced by $x_M$. In other words,
the a.c. susceptibility measurements are only sensitive to the deepest level of
the tree. Actually, Eq. (33-b) for $\chi''$ is very well obeyed experimentally
-- although a `non aging' contribution $\chi''_{\rm eq.}$ must be added: see
Fig. 4. Indeed, in the limit $\overline \omega \longrightarrow \infty$, the
short time dynamics corresponding to intra-state fluctuations ($x > 1$) cannot
be neglected in Eq. (33).\par
As mentionned above, magnetic noise measurements have also been performed in
order to test FDT in spin-glasses \cite{Ocio}. The `usual' FDT would predict
that the noise spectrum $S(\omega,t)$ is related to $\chi''(\omega,t)$ through:
$$
\chi''(\omega,t) = {\pi X \over 2} {\omega S(\omega,t) \over T} \eqno(34)
$$
with $X \equiv 1$. The non-equilibrium theory proposed in
\cite{CuKu},\cite{CuKu2},\cite{MF} claims that $X = x_{M} < 1$ in the limit
$\omega t \gg 1$, while our model suggests (in the same limit) $X = 1 - {\zeta
\over \omega t}$ -- see also \cite{Koper2}. Experimentally, the noise
measurements were not calibrated in an absolute way. The proportionality
relation (34) was indeed confirmed but no value of $X$ could be determined. In
view of the present discussion, it seems to us that this would be a very
important point to check: the value of $x_M$ extracted from the decay of
$\chi''$ or the TRM should be the same as the proportionality constant
extracted from Eq. (34). Since $x_M$ is in the range $0.6-0.9$, a significant
effect could show up by comparing carefully equilibrium and non-equilibrium
regimes.

Let us finally note that for $x > 1$, $\chi''(\omega,t)$ is given by:
$$\chi''(\omega,t) \propto \omega \tau_0^{x-1} t^{2-x}\qquad {\rm for}\  \omega
\ll t^{-1} \eqno(35-a)$$
$$
\chi''(\omega,t) \propto (\omega \tau_0)^{x-1} \qquad {\rm for}\  t^{-1} \ll
\omega \ll \tau_0^{-1} \eqno(35-b)
$$
(the large frequency region $\omega \gg \tau_0^{-1}$ requires a detailed
description of $p(\tau)$ for $\tau \ll \tau_0$).
Hence for $x \simeq 1$ (i.e slightly above the spin glass transition, or
equilibrium dynamics in the spin-glass phase), Eqs (33-35) gives $S(\omega)
\propto {T \over \omega}$ i.e. the model naturally leads to $1/f$ noise.

\subsection {Temperature cycling}

As mentioned in the introduction, more sophisticated experimental protocoles
have been investigated \cite{exp3},\cite{exp4}, where the temperature or
magnetic field is changed {\it during} TRM or $\chi''$ measurements. A complete
analysis of these experiments is beyond the aim of the present paper and will
be given in separate publications \cite{ustocome}. We shall restrict here to a
qualitative discussion of the results, focusing on the remarkable memory effect
mentionned in the introduction, and illustrated in Fig. 5 (from \cite{figT}).

Let us thus discuss the effect of a small temperature change on $\chi''$ within
a REM landscape. As discussed in the previous section, this REM description is
sufficient to account for the behaviour of $\chi''(\omega,t)$ in the limit
$\omega t \gg 1$ (provided an equilibrium contribution is added). In the REM,
the (free-) energy landscape does not evolve with temperature and $\overline x$
is simply given by $T \over T_g$. A small change of temperature $T
\longrightarrow T'=T-\Delta T$ simply changes the trapping times as: $\tau
\longrightarrow \tau'= \tau_0 ({\tau \over \tau_0})^p$ with $p={T \over T'} >
1$. Suppose that the waiting time before the small temperature jump is $t_{w1}$
-- see Fig 5. As emphasized in \cite{JPB} and above, only traps with $\tau
\simeq t_{w1}$ have a significant probability to be observed. Hence, if the
time spent at temperature $T'$ is such that $t_{w2} \ll  \tau_0
({t_{w1} \over \tau_0})^p$, it is quite obvious that no jump will take place
during this intermediate period. The `memory effect'
$\chi''(\omega,t_{w1}+t_{w2}+0)=\chi''(\omega,t_{w1}-0)$ would then be trivial.
On the other hand, one can show that the change of $\chi''$ at $t_{w1}$ is
given by:
$$
{\chi''(\omega,t_{w1}+0) \over \chi''(\omega,t_{w1}-0)} = p
{q_{EA}(T')X(T')\over  q_{EA}(T)X(T)} (\omega \tau_0)^{(\overline x-1)({1\over
p} -1)} \eqno(36)
$$
Typically (see Fig 5), $p = 1.2$, $\overline x = .75$, while the variation of
$q_{EA}$ is roughly a factor $1.2$. Hence, provided $\tau_0 < 10^{-4}$ sec.,
$\chi''$ should first decrease and then remain constant when the temperature
jumps down, at variance with the experimental results. In physical terms, all
dynamical processes should be slowed down by the temperature jump -- hence the
decrease of $\chi''$ and the memory effect. The {\it simultaneous} observation
of a strong increase of $\chi''$ and $\left|{\partial \chi'' \over \partial
t}\right|$ (`rejuvenation') and of the memory effect is non trivial
\cite{Drop?}. In \cite{exp3}, it was suggested that the origin of this combined
effect comes from the hierarchical nature of the phase-space, with valleys
bifurcating into subvalleys at all temperatures below $T_g$ in a continuous
sequence of `micro phase transitions'. Even a small temperature change can thus
be thought as a `quench' from `high' temperature, setting back the age of the
system to zero -- hence the increase of $\chi''$.

We want to rephrase this interpretation within the framework developed here. As
mentionned above, the construction of Parisi's tree can be extended `within
states' to $x > 1$ and $q > q_{EA}$, simply reflecting the fact that even
within each state there are long-lived configurations and non exponential
equilibrium dynamics, even above $T_g$. However, as mentionned previously, only
the part of the tree corresponding to $x < 1$ corresponds to non-stationary
dynamics, while the part $x > 1$ is equilibrated within microscopic time
scales. $T_g$ is the temperature at which the smallest $x$ first reaches the
critical value $1$, but one can expect a whole sequence of phase transitions as
the temperature is decreased, corresponding to the crossing of $x = 1$ for the
successive levels of the tree, much as in the `generalized' Random Energy Model
\cite{GREM}.
(As discussed below, this could correspond in real space to the progressive
freezing of the dynamics over smaller and smaller length scales.)
Hence if at a certain temperature $x_{M} < 1$ but $x_{M+1} > 1$, only the
transitions at the $M^{\rm th}$ level of the tree will contribute to the
relaxation of $\chi''$. But if upon lowering the temperature $x_{M+1}$ becomes
less than $1$, then the processes corresponding to this level of the tree will
start contributing, while those corresponding to $x_{M}$ are to a certain
extent frozen -- allowing for the memory effect. The small temperature jump is
furthermore tantamount to a {\it quench} for the $M+1$ level since when
$x_{M+1} > 1$, all states are more or less equivalent, corresponding to a
random initial condition at $t=t_{w1}+0$.

Another related scenario is also possible: as discussed in section II.3, the
finiteness of the ergodic time  $\te^j$ at level $j$ leads to `interrupted
aging' \cite{JPB}, after a time $\te^j$. The crucial levels of the tree are in
this case not those such that $x_{M} <  1 < x_{M+1}$ but, as discussed in
section II.3, those such that $\te^M(T) \gg t_w \gg \te^{M+1}(T)$, even if
$x_{M+1} < 1$. Aging processes will be restarted upon cooling if
$\te^{M+1}(T-\Delta T) \gg t_w$.
A quantitative analysis of the experimental data along these lines is currently
underway. Experiments investigating the role of the magnetic field on aging and
their interpretation are reported in \cite{deltah}.

\subsection{Noise second spectrum}

Another interesting set of experiments which can be discussed within our model
is the noise `second spectrum' measurements of Weissmann et al. \cite{Weiss}.
Very briefly, the point is that on sufficiently small samples, one observes $1
\over f$ noise, but with an amplitude which is itself `noisy', i.e. randomly
changing with time (albeit on rather long time scales.) The natural
interpretation -- discussed in different terms in \cite{Weiss} -- is that the
noise primarily comes from the near equilibrium levels $x \simeq 1$, but that
higher level jumps will slightly change the amplitude of this noise -- since
the systems are small enough, the averaging over different `branches' is not
performed. For a multilayer tree, the prediction is obviously that the noise
spectrum behaves as $\omega^{x_M -2} \simeq \omega^{-1}$, but with a {\it noisy
amplitude} with a correlation function decaying as $t^{-x_1}$, as observed
experimentally (where this exponent is called $\beta$). A more quantitative
analysis of the experimental results along these lines would however be
desirable, but we note that the value of $\beta=x_1$ determined for CuMn in
\cite{Weiss} is very close to the one quoted in Table 1,2 for $x_1$, and
evolves similarly with temperature.

\section{DISCUSSION. TREES AND DROPLETS}

Fisher and Huse \cite{FH}, see also \cite{KH} were the first to stress the
importance of understanding the nature of the spin-glass excitations in finite
dimensional spaces. They argued that these excitations should be of the form of
{\it compact} `droplets', with a surface much smaller than their volume. These
droplets were defined with respect to the (unique up to a global spin-flip)
ground state as the `spin-flip' excitation of lowest possible energy within a
region of size $L$; the energy of such a droplet is found to be of order
$L^\theta$ ($\theta \simeq 0.2$ in $d=3$), and the energy barrier for such a
droplet to be activated grows as $L^\psi$. Apart from the presence of these
ever-growing energy barriers, the description of Fisher and Huse of a
spin-glass is that of a `disguised ferromagnet'. In particular, exactly as in a
ferromagnet quenched in zero field, a spin-glass quenched below $T_g$ would
approach equilibrium by growing domains of the two `pure phases' assumed to be
present, thereby eliminating the extra energy ($L^\theta$) associated with the
domain walls. The system thus {\it coarsens} with time; after time $t_w$ the
typical size of the domains is $R(t_w)=\log^{1\over \psi}({t_w \over \tau_0})$
-- instead of $R(t_w)=t_w^{1/2}$ in a ferromagnet (with non conserved dynamics,
see \cite{Bray}). Before discussing the form of the relaxation proposed in
\cite{FH} inspired from this picture, it is useful to recall how $C(\tw,t_w)$
would look like for a simple ferromagnet. An exactly soluble case is the
$\phi^4_n$ theory in the limit $n \longrightarrow \infty$, for which one finds
\cite{Bray}:
$$
C_{n = \infty}(\tw,t_w) = \lb {4 (\tw)t_w \over (2t_w + t)^2} \rb^{d \over
4}\eqno(37)
$$
(d is the dimension of space). Mazenko's approximate theory for $n=1$
\cite{CuKuPa},\cite{Bray} leads to:
$$
C_{n=1}(\tw,t_w) = {4 \over \pi} \tan^{-1}\sqrt{2-C_{n = \infty}(\tw,t_w) \over
C_{n = \infty}(\tw,t_w)}\ -1 \eqno(38)
$$
For $t \gg t_w$, expressions (37,38) simplify to $({t_w \over t})^{d/4} \equiv
({R(t_w) \over R(t)})^\lambda$ with $\lambda={d \over 2}$, whereas more
elaborate theories show that $\lambda$ is non trivial in general \cite{Bray}.
Fisher and Huse argued that the appearance of the ratio of two length scales,
$R(t)$ and $R(t_w)$, should be general and proposed for the spin-glass dynamics
to write:
$$
C_{FH}(\tw,t_w) \propto R(t)^{-\theta} \Sigma({R(t) \over R(t_w)})\eqno(39)
$$
with $\Sigma(x \longrightarrow 0) = {\rm const.}$ and $\Sigma(x \longrightarrow
\infty) = x^{\theta-\lambda}$. (Again, we identify $C(\tw,t_w)$ and
$M(\tw,t_w)$ in the small $H$ limit). Eq. (39) means that:\par
a) The scaling variable for aging should be $\log({t \over \tau_0})\over
\log({t_w \over \tau_0})$ rather than ${t \over t_w}$.
\par
b) Even in the limit where $t_w = \infty$, $C_{FH}(\tw,t_w)$ decays to zero at
large times if $\theta >0$, at variance with Eq. (11-b) and the results of
\cite{CuKu},\cite{CuKu2},\cite{MF}.

Before comparing with the experimental data, let us remark that, from  Eqs.
(37-38), $C(\tw,t_w)$ is a function of $t \over t_w$ {\it over the whole range}
of time scales. This scaling variable is thus more general than the asymptotic
one, i.e. ${R(t) \over R(t_w)}$. This suggests that even in the case of
logarithmic domain growth, a ${t \over t_w}$ scaling should hold for
coarsening. We have confirmed this \cite{ustocome2} for the 1d Random Field
Ising Model where $R(t) \propto \log^4 t$ (see e.g.
\cite{PhysRep},\cite{PaMa}). As emphasized in \cite{CuKuPa}, Eqs. (37-38) show
that aging is not specific of spin-glasses, but occur as soon as the
equilibration time of the system is infinite (or much larger than $t,t_w$). We
have plotted in Fig. 6 $C_n(\tw,t_w)$ as given by Eqs. (37-38) as a function of
$t \over \tw$. The major difference with the spin-glass  data is the singular
behaviour of the latter for short times (see Figs. 2-a,b).

We have tried to test the predictions contained in Eq.(39). For example, we
have looked for the best values of $\theta$ and $\tau_0$ to rescale TRMs at a
fixed temperature and different $t_w$, for both AgMn and CrIn. The data points
unambiguously towards $\theta=0$; data collapse is quite good at short times
but deteriorates as soon as $t > t_w$ : see Fig. 7. The other test is to
compare different temperatures with the same $t_w$. We fixed $\theta=0$, and
chose $\tau_0(T)$ and the T-dependent proportionality constant in Eq. (39)
(which is essentially $q_{EA}(T)$) to best rescale the {\it late} part of the
curves. Although this late part can indeed be satisfactorily fitted by
$\log^{-{\lambda\over \psi}} ({t\over \tau_0(T)})$ with ${\lambda\over \psi}
\simeq 1.2$, the early part of the data does not scale (Fig. 8). Hence we
believe that Eq. (39), even with $\theta=0$, is inadequate to describe the data
consistently.

Koper and Hilhorst \cite{KH} proposed a similar picture, although they assumed
that $R(t)$ grows as a power-law $t^p$ rather than logarithmically. Also, the
relaxation time associated to a domain of size $R$ is taken to be $\propto
R^z$.
It would be too long to discuss in detail this theory here; let us simply
mention that it does lead a relaxation of $\chi''(\omega,t)$ behaving as
$\omega^{-1} t^{-pz}$. Consistency with the $\omega t$ (or $t \over t_w$)
scaling observed experimentally leads to $pz \simeq 1$ -- which is in fact
natural since it means that the growth time and relaxation time of droplets are
comparable \cite{Koper3}. However, the resulting relaxation of $\chi''$ as
$t^{-1}$ is much too fast (compare with Eq. (34), with $\overline x \simeq
.75$).

The basic remark of Fisher, Huse, Koper and Hilhorst that the flipping spins
are clustered somewhere in space and that the time scales should grow with the
size of these clusters seems however unavoidable. How can this be reconciled
with replica symmetry breaking ? Of course, a proper replica (or dynamical)
theory in finite dimension is needed to answer completely this question. Such a
theory is not yet available for spin glasses, but has been worked out for the
simpler problems of manifolds in random media \cite{RM}, \cite{BMY} (i.e.
polymers, surfaces, vortex lattices. etc..). One basically finds that $q(x)$
becomes $L$ (scale) dependent, with a characteristic value of $x(L)$ varying as
$L^{-\theta}$: small scales correspond to large $x$.

{}From Eq. (1) and the interpretation of $x(L)$, one thus sees that the energy
distribution for the excitations (`droplets') of scale $L$ has the form:
$$
\rho(f,L) = {1 \over f_0 L^{\theta}} \exp(-{|f|\over f_0 L^\theta })\eqno(40)
$$
showing, as postulated by Fisher and Huse, that the energy scale grows as
$L^\theta$. At this stage, an important difference with Fisher and Huse is the
exponential form of the distribution. Interestingly, Eq. (40) means, within our
interpretation, that there is a spin-glass transition temperature $T_g(L)$
associated with each scale, defined through $x(L,T_g(L))=1$. The infinite
sequence of micro-phase transitions suggested in \cite{exp2},\cite{exp3} thus
corresponds to a progressive `weak ergodicity breaking' (in the sense that
$x(L)$ crosses 1, see II.1) of smaller and smaller length scales (faster and
faster degrees of freedom).

Quite naturally, one can associate an L-dependent overlap $q(L)$ between
configurations obtained by flipping a droplet as $1 - q(L) \propto ({L \over
\xi})^{d_f}$, where $d_f$ is the fractal dimension of the droplets ($d_f$ is
equal to $d$ for Fisher and Huse) and $\xi$ is a correlation length. Hence, the
picture we propose is in fact very close the original droplet model, except
that: \par
a) The `droplets' are only required to be metastable, and not `lowest'
excitations, \par
b) These droplets may a priori be non compact ($d_f < d$), and\par
c) The time scale associated with droplets of size $L$ {\it is not peaked}
around some $\tau(L)$, but rather {\it power-law distributed} with a parameter
$x(L)$ which becomes smaller and smaller at large sizes.\par
d) The Fisher-Huse time scale $\tau_0 \exp(L^\theta)$ could be interpreted as
the ergodic time $\te(L)$ associated to scale $L$: one indeed should expect
that Eq. (40) ceases to be valid for $|f| >>>  f_0 L^\theta$.\par
e) Aging will be totally interrupted when the 'terminal' ergodic time scale
$\te(\xi)$ associated to $\xi$ is reached.\par

This picture suggests that the final description of the experiments should
involve a continuous tree, rather than the 2-level approximation that we have
chosen. Further work on this aspect and on the related problem of the
Fluctuation Dissipation theorem is certainly needed; we however hope that the
scenario proposed here can be helpful to think about non ergodic dynamics in
glassy systems, in particular those in which quenched disorder is a priori
absent \cite{DREM}. We note in this respect that the spontaneous appearance of
power-law distributed `trapping' times with $x$ crossing 1 at the glass
transition has been reported for hard-sphere systems \cite{japs}.

{\it Acknowledgments} We wish to express warm gratitude towards A. Barrat, L.
Cugliandolo, S. Franz, J. Hammann, M. M\'ezard, M. Ocio, E. Vincent, M.
Weissmann for enlightning critical discussions. We also take this opportunity
to acknowledge the inspiring work of L. Cugliandolo, S. Franz, J. Kurchan and
M. M\'ezard on this problem.

\vskip 0.5cm
{\it Figure Captions}
\noindent\par {\bf Figure 1} Schematic views of the free energy landscape with
associated Parisi trees. 1-a: REM landscape, 1-b: Full RSB landscape.
\noindent\par {\bf Figure 2} 2-a : Fit of TRM decay for CrIn at 10 K against
two level tree theory with simple FDT. Note that the fit is perfect over the
whole time domain. 2-b : Fit of TRM decay for AgMn at 9 K against two level
tree theory with simple FDT.
\noindent\par {\bf Figure 3} Fit of TRM decay for AgMn at 9 K against two level
tree theory with generalized FDT as given by equation (26).
\noindent\par {\bf Figure 4} Fit of the decay of the out of phase
susceptibility of CrIn (at T=12 K, $\omega=10^{-2}$ Hz.) as $\chi''(\omega,t) -
\chi_{\rm eq.}(\omega) \propto t^{x_2 -1}$, and $x_2 = .79$ (see Table 2).
\noindent\par {\bf Figure 5} Sketch of the temperature cycling protocole and
the evolution of $\chi''$ during this cycling (from \cite{figT}). Note the {\it
strong spike} just after the temperature decrease followed by a {\it perfect
`memory'}  in the third stage.
\noindent\par {\bf Figure 6} Aging in $n=1$ (approximate result) and $n
=\infty$ (exact result) ferromagnets. Note that the small time behaviour is
regular, at variance with the experimental data on spin glasses (Figs 2-a,
2-b).
\noindent\par {\bf Figure 7} $M(\tw,t_w) \log^{\theta/\psi}({t \over \tau})$ as
a function of $\log({t \over \tau})\over
\log({t_w \over \tau})$ for AgMn at 8K. The values of $\tau=10^{-5}$ sec. and
${\theta \over \psi } =0$ have been chosen to obtain the best rescaling.
\noindent\par {\bf Figure 8}  ${M(\tw,t_w) \over q_{EA}(T)} $ as a function of
$\log({t \over \tau(T)})\over
\log({t_w \over \tau(T)})$ for CrIn at various temperatures. The values of
$\tau(T)$ sec. and $ q_{EA}(T)$ have been chosen to obtain the best rescaling
of the late part of the curve, from which we extract ${\lambda \over \psi }
\simeq 1.2$.
\noindent\par{\bf Tables} Extracted values of $x_1$, $x_2$, $q_{EA}$ and $q_1$,
from CrMn and AgMn TRM decay data (using simple form of the FDT).

\centerline{Appendix A - Some properties of exponential times}

Here we shall review and give  proofs of some of the properties of exponential
times used in this paper.
 The probability density function of an exponential random variable $T$ of rate
$E$, is given by
$$ p(t) = E\exp(-Et), \qquad t\geq 0 .$$
\par \noindent{\bf Property A-1:}
The exponential time is {\it memoryless}, that is conditional on the time $T$
not having occurred at some time $t$, the distribution of the subsequent time
$T^\prime$ before it occurs is the same as that for $T$.
 The proof uses  Bayes' theorem to show that, the conditional probability
density function for $T^\prime$ is
$$ \rho(t^\prime )  = {E\exp\lb E(t + t^\prime) \rb \over \int_t ^\infty
dsE \exp(-Es) } = E\exp(-Et^\prime), $$
and hence the result.
\par \noindent {\bf Property A-2:}
Given two independent exponential times $T$ and $T^\prime$ of rates $E$ and
$E^\prime$ respectively, their minimum is distributed as an exponential random
time of rate $E + E^\prime$. This can be seen as follows: if the probability
density function for $\min(T,T^\prime )$ is $\rho (u)$, then
$$ \rho (u) = \int dtdt^\prime E E^\prime \exp(-Et -E^\prime t^\prime )\delta
\lb u - \min(t,t^\prime) \rb = (E + E^\prime)\exp\lb -(E + E^\prime) u\rb.$$
By induction we see that for any finite number of independent exponential times
the minimun is distributed as an exponential with rate given by the sum of
their rates.
\par\noindent{\bf Property A-3:}
Given two independent exponential times $T$ and $T^\prime$ of rates $E$ and
$E^\prime$ respectively then
$$ P(T< T^\prime) = \int_{t<t^\prime} dtdt^\prime E E^\prime \exp(-Et -E^\prime
t^\prime ) = {E\over E + E^\prime} .$$
This property generalises to: for $j$ independent exponential times $T_i \quad
1\leq i\leq j$ of rates $E_i \quad 1\leq i\leq j$ respectively, then
$$ P(T_k = \min\{T_j, \quad 1\leq i\leq j\}) = {E_k\over \sum_{i=1}^j E_i}.$$
The proof is simply uses the basic form of property (A-3) and also property
(A-2).

\bibliographystyle{unsrt}

\begin{thebibliography}{99}
\bibitem{Silvio} We are indebted to Silvio Franz for an important discussion on
this point.
\bibitem{longtime} The late time power law behaviour $t^{-x_1}$ can be probed
on nearly four decades using the attenuation of aging induced by a strong
magnetic field : see \cite{deltah}
\bibitem{Koper3} In a latter paper \cite{Koper2} the authors study analytically
the 1d $\pm J$ spin glass. They indeed find $p \equiv {1 \over z} = {1 \over
2}$, and compute  $M(\tw,t_w)$. Interestingly, their result can be written {\it
exactly} as in Eq. (9) with $x = {1\over 2}$. This is not a coincidence: the
probability for a domain to survive during a time $\tau$ is given by the
probability for two random walkers (the domain walls) to meet for the first
time after time $\tau$, which is, in 1d and for large $\tau$, $\propto
\tau^{-1-{1 \over 2}}$.
\end{thebibliography}

\end{document}